\documentclass[a4paper, amsfonts, amssymb, amsmath, reprint, showkeys, nofootinbib, twoside,notitlepage,onecolumn]{revtex4-1}

\usepackage{amsmath,amstext}
\usepackage[T1]{fontenc}
\usepackage{amssymb}
\usepackage{graphicx}
\usepackage{ae,aecompl}

\DeclareFontFamily{OT1}{pzc}{}
\DeclareFontShape{OT1}{pzc}{m}{it}{<-> s * [1.10] pzcmi7t}{}
\DeclareMathAlphabet{\mathpzc}{OT1}{pzc}{m}{it}

\usepackage{hyperref}
\usepackage{amsmath}
\usepackage{amssymb}
\usepackage{mathtools}
\usepackage{bm}
\usepackage{cleveref}
\usepackage{tensor}
\usepackage{braket}
\usepackage{enumitem}
\usepackage{mhchem}
\usepackage{amsthm}
\usepackage{nccmath}
\usepackage{mathrsfs}
\usepackage{color}

\newcommand{\feq}{{f_{\mathrm{eq}}}}

\def\be{\begin{equation}}
\def\ee{\end{equation}}
\def\beq{\begin{eqnarray}}
\def\eeq{\end{eqnarray}}

\theoremstyle{definition}

\theoremstyle{theorem}

\theoremstyle{corollary}

\begin{document}
\title{The quasi-normal modes of relativistic Fokker-Planck kinetic theory}
\author{L.~Gavassino}
\affiliation{Department of Applied Mathematics and Theoretical Physics, University of Cambridge, Wilberforce Road, Cambridge CB3 0WA, United Kingdom}

\begin{abstract}
Employing the well-known unitary equivalence between Fokker-Planck operators and Schr\"odinger Hamiltonians, we compute the quasi-normal-mode spectrum of ultrarelativistic kinetic theories with momentum-space diffusion. We show that the collision operator reduces to a Dirac-delta Schr\"odinger problem in one spatial dimension, and to a Coulomb Schr\"odinger operator with hydrogenic spectrum in three dimensions. Finite spatial wavenumber appears as a perturbation of the associated quantum potential. The hydrodynamic mode is found to obey exact Fick-type diffusion at all real wavenumbers, whereas relativistic kinematics generically produces a continuous ballistic band in the non-hydrodynamic sector, a feature absent in the Newtonian regime.
\end{abstract}

\maketitle

\section{Introduction}

Quasi-normal modes furnish a concise spectral description of the linear response properties of a dissipative system.
Within hydrodynamics and kinetic theory, they are defined as all the solutions of the linearized evolution equations with a spacetime dependence of the form \(e^{i k x - i \omega t}\), for a fixed spatial wavenumber \(k\)\footnote{If the number of spatial dimensions is greater than 1, one should in principle deal with wavevectors $k^j$. In practice, however, one can fix the direction of propagation, and align the $x$-axis with the wavevector so that $k^j=(k,0,0,...)$.}, and with eigenfrequency \(\omega(k)\).
For real \(k\), the real part of \(\omega(k)\) usually characterizes the oscillatory and propagating aspects of the response, while the imaginary part typically controls the rate of equilibration and spatial spreading.
Owing to this interpretation, quasi-normal modes have emerged as a key framework for elucidating the onset of hydrodynamic behavior
\cite{Dudynski1989Hydro,Grozdanov:2019kge,GavassinoConvergence:2024xwf,GavassinoDisperisonRadiation2024cqw},
as well as for characterizing transient dynamics beyond the hydrodynamic regime
\cite{KovtunHolography2005,Denicol_Relaxation_2011,RomatschkeCutsandPoles:2015gic,Kurkela:2017xis,Moore:2018mma,GavassinoGapless:2024rck,Bajec:2024jez,Brants:2024wrx}.

The quasi-normal-mode spectra of relativistic kinetic theories have been computed for a broad class of collision integrals.
These include the Relaxation-Time Approximation (RTA)
\cite{RomatschkeCutsandPoles:2015gic,Bajec:2024jez,Brants:2024wrx},
weakly coupled \(\lambda\phi^4\) kinetic theory
\cite{Moore:2018mma,RochaBranchcut:2024cge,GavassinoGapless:2024rck}, generic Boltzmann-type integrals \cite{DudynskiEkielJezewska1985,Dudynski1989Hydro},
and radiative transfer models
\cite{GavassinoDisperisonRadiation2024cqw,GavassinoRadiativeBounds:2025bxx}.
There is, however, a further class of collision integrals that possess a particularly tractable analytical structure and play a key role in descriptions of early-time gluon dynamics in heavy-ion collisions
\cite{Mueller:1999pi,Blaizot:2013lga}:
Fokker-Planck kinetic theories.
The objective of this work is to determine the quasi-normal-mode spectrum of a minimal representative of such theories in the ultrarelativistic regime, for arbitrary \(k\in \mathbb{C}\).

Our guiding principle is the well-known fact that detailed-balanced Fokker-Planck generators are unitarily equivalent to Schr\"odinger-type operators acting on an $L^{2}$ Hilbert space. This correspondence underlies much of the modern spectral analysis of Fokker-Planck equations and plays a central role in hypocoercivity theory and the study of relaxation to equilibrium \cite{RiskenFP,VillaniHypocoercivity,HelfferNier,PavliotisBook}. Within this framework, the kinetic equilibrium is identified with the ground state of the associated Schr\"odinger operator, while relaxation rates are determined by its excitation spectrum.
In the nonrelativistic setting, the canonical example is provided by the Ornstein-Uhlenbeck operator, whose ground-state transform yields a shifted quantum harmonic oscillator (we will review this result in Section \ref{notablepotentials}). This equivalence has been extensively exploited to obtain exact spectra and decay rates, and it provides a paradigmatic illustration of how quasi-normal modes arise from the discrete excitation spectrum of the dual Schr\"odinger problem \cite{RiskenFP,PavliotisBook}. The harmonic-oscillator structure has also served as a benchmark for more general hypocoercive kinetic models.

There exist analogous Fokker-Planck models also within relativistic kinetic theory, with the equilibrium state given by the Maxwell-J\"uttner distribution. Such systems have been investigated both from the probabilistic perspective of relativistic Brownian motion and from the standpoint of kinetic theory \cite{Debbasch,DunkelHanggi,Felderhof2012}. However, despite this progress, the quasi-normal-mode spectrum has never been computed in detail.
In this work, we will apply the unitary equivalence to show that the quasi-normal modes of ultrarelativistic Fokker-Planck kinetic theory are exact solutions of the time-independent Schr\"odinger equation with a perturbed potential (where $k$ is the perturbation parameter). We will find that the structure and physical interpretation of such a potential change depending on the number of spatial dimensions, and so does the resulting spectrum. All in all, the results presented here will place ultrarelativistic Fokker-Planck kinetics on the same conceptual footing as the well-known harmonic-oscillator structure of nonrelativistic Ornstein-Uhlenbeck dynamics.

Throughout the article, we work in a Minkowski $(D+1)$--dimensional spacetime, with metric signature $(-,+,+,...)$, and adopt natural units, $c=\hbar=k_B=1$. Greek indices $\mu,\nu$ are spacetime indices (they run from $0$ to $D$), while $j,k$ are pure spatial indices (they run from $1$ to $D$).  

\section{The Vlasov-Fokker-Planck equation: A quick overview}
\vspace{-0.1cm}

In this section, we review the basics of Fokker-Planck kinetic theory for a nondegenerate system of (quasi)particles, with particular emphasis on its mathematical structure and properties.

\subsection{The linearized Boltzmann equation}

We consider a dilute gas of (quasi)particles propagating through a medium at rest. The (quasi)particles undergo elastic and inelastic scattering off impurities in the medium, so that neither the total energy nor the total momentum of the gas is conserved. We assume, however, that particle creation and annihilation are forbidden, ensuring conservation of the total (quasi)particle number.

Let \(\varepsilon=\varepsilon(p_j)\) denote the dispersion relation of the (quasi)particles, where \(\varepsilon\) is the single-particle energy and \(p_j\) the momentum. We denote by \(f(t,x^j,p_j)\) the kinetic distribution function of the gas. In equilibrium, the distribution function takes the form \(\feq=e^{\alpha-\beta \varepsilon}\), where \(\beta\) is the inverse temperature \textit{of the medium}. The parameter \(\alpha\) is not fixed a priori, and is determined by the total (quasi)particle number of the gas. Now, fix an equilibrium state, and write $f{=}\feq{+}\Phi$, where $\Phi(t,x^j,p_j)$ is a small perturbation. Then, the linearized Boltzmann equation reads \cite[\S 74]{landau10}
\begin{equation}\label{BoltzmannEquation}
(\partial_t+v^j \partial_j)\Phi=-I\Phi\, ,
\end{equation}
where $v^j=\partial \varepsilon/\partial p_j$ is the (quasi)particle velocity, and $I$ is some linearized collision operator. 

\subsection{Properties of the linearized collision operator}

Fix a spacetime point \((t,x^j)\), and consider perturbations \(\Phi(p_j)\) belonging to the Hilbert space
\(
\mathfrak{H}=L^2\!\left(\mathbb{R}^D,\frac{d^Dp}{(2\pi)^D f_{\mathrm{eq}}}\right)
\)
defined at that point. The associated inner product is
\begin{equation}\label{inner product}
(\Phi,\Psi)
=\int_{\mathbb{R}^D}\frac{d^Dp}{(2\pi)^D f_{\mathrm{eq}}}\,
\Phi^*\,\Psi \, .
\end{equation}
As shown in Appendix \ref{aaa}, the collision operator \(I\), regarded as a linear operator on \(\mathfrak{H}\), must always satisfy the following physical consistency conditions \cite{cercignani_book,GavassinoGapless:2024rck,GavassinoConvergence:2024xwf,GavassinoDistrubingMoving:2026klp} (for all
\(\Phi,\Psi\in\mathfrak{H}\)):
\begin{equation}\label{Properties}
\begin{aligned}
&(\Phi,I\Phi)\ge 0 ,\\
&(\Phi,I\Psi)=(I\Phi,\Psi) ,\\
&I\Phi=0 \quad \text{if and only if} \quad
\Phi=\mathrm{const}\times f_{\mathrm{eq}} .
\end{aligned}
\end{equation}
In particular, \(I\) is a self-adjoint, non-negative operator with
\(\ker(I)=\mathrm{span}\{f_{\mathrm{eq}}\}\).

The assumption on the kernel of the collision operator has direct implications for the admissible equilibrium states of the theory. Defining an equilibrium configuration \(\Phi_{\mathrm{eq}}\) as a state that is homogeneous and stationary, one finds from equation \eqref{BoltzmannEquation} that the only such solutions are proportional to \(f_{\mathrm{eq}}\), i.e.\ \(\Phi_{\mathrm{eq}}=a\,f_{\mathrm{eq}}\) with \(a=\mathrm{const}\). The corresponding distribution functions take the form
\begin{equation}\label{equilibrazione}
f=f_{\mathrm{eq}}+a f_{\mathrm{eq}}=(1+a)f_{\mathrm{eq}}
\overset{\mathcal{O}(a)}{=} e^{a} f_{\mathrm{eq}}
= e^{(\alpha+a)-\beta \varepsilon}\, ,
\end{equation}
which represent equilibrium states with different total (quasi)particle numbers, corresponding to a shift of the parameter \(\alpha\). Equilibrium states with different temperatures or nonzero bulk velocities are not admitted, since the medium acts as a thermal bath; this is reflected in the lack of energy and momentum conservation for the gas alone.

\subsection{The Fokker-Planck ansatz}

Fokker-Planck kinetic theory corresponds to a particular realization of the Boltzmann equation~\eqref{BoltzmannEquation}, in which the collision operator takes the form \cite{Debbasch,DunkelHanggi}
\begin{equation}\label{fokkerplanck}
I\Phi=-\nu\,\frac{\partial}{\partial p^j}
\left(\frac{\partial \Phi}{\partial p_j}
+\beta v^j \Phi \right) \, ,
\end{equation}
with \(\nu>0\) a constant parameter (= ``momentum diffusivity''). This operator models the cumulative effect of many uncorrelated small-angle scattering events, leading to diffusive dynamics in momentum space. The associated drift term ensures relaxation toward the equilibrium distribution~\eqref{equilibrazione} at late times.
\newpage

One readily verifies that this collision operator satisfies all properties in \eqref{Properties}. Using \(f_{\mathrm{eq}}^{-1}=e^{-\alpha+\beta \varepsilon}\), one obtains
\(
\partial_{p_j}\!\left(f_{\mathrm{eq}}^{-1}\Phi\right)
= f_{\mathrm{eq}}^{-1}\!\left(\partial_{p_j}\Phi+\beta v^j\Phi\right)
\),
which allows the collision operator to be recast in the equivalent form
\begin{equation}\label{III}
I\Phi=-\nu\,\frac{\partial}{\partial p^j}
\left[
f_{\mathrm{eq}}\,
\frac{\partial}{\partial p_j}
\left(f_{\mathrm{eq}}^{-1}\Phi\right)
\right] \, .
\end{equation}
Substituting \eqref{III} into the inner product $(\Phi,I\Psi)$, and integrating by parts, yields
\begin{equation}
(\Phi,I\Psi)
=\nu\int_{\mathbb{R}^D}\frac{d^Dp}{(2\pi)^D}\,
f_{\mathrm{eq}}\,
\frac{\partial}{\partial p^j}
\left(f_{\mathrm{eq}}^{-1}\Phi^*\right)\,
\frac{\partial}{\partial p_j}
\left(f_{\mathrm{eq}}^{-1}\Psi\right) \, ,
\end{equation}
which makes the self-adjointness and non-negativity of \(I\) manifest. Moreover, this expression shows that \((\Phi,I\Psi)=0\) for all \(\Phi\) (and hence \(I\Psi=0\)) if and only if
\(\partial_{p_j}(f_{\mathrm{eq}}^{-1}\Psi)=0\), i.e.\ \(\Psi=\mathrm{const}\times f_{\mathrm{eq}}\), as required.

\section{Mathematical correspondence with the Schr\"odinger equation}

We now demonstrate that the quasi-normal modes of the Boltzmann equation~\eqref{BoltzmannEquation} with the Fokker-Planck collision operator~\eqref{fokkerplanck} can be formulated as solutions of an eigenvalue problem for a perturbed Schr\"odinger operator. Although this kind of transformation is standard in the analysis of Fokker-Planck dynamics \cite[\S 5.4, \S 5.5]{RiskenFP}, its application to the computation of the quasi-normal spectrum in kinetic theory appears to be new.

\subsection{Derivation of the correspondence}

We seek solutions of the Boltzmann equation of the form
\(
\Phi(t,x^j,p_j)=\Phi(p_j)\,e^{ikx^1-i\omega t}
\),
with \(k,\omega\in\mathbb{C}\). For this spacetime dependence, equation \eqref{BoltzmannEquation} reduces to
\begin{equation}\label{eigenmode}
\left(-E+\chi v^1\right)\Phi
=\frac{1}{2}\,\frac{\partial}{\partial p^j}
\left[
f_{\mathrm{eq}}\,
\frac{\partial}{\partial p_j}
\left(f_{\mathrm{eq}}^{-1}\Phi\right)
\right]\, ,
\end{equation}
where we have introduced the rescaled parameters
\(E=i\omega/(2\nu)\) and \(\chi=ik/(2\nu)\). 

Now, we perform the transformation
\(
\Phi=\sqrt{(2\pi)^D f_{\mathrm{eq}}}\,\psi
\),
under which the inner product~\eqref{inner product} reduces to the standard \(L^2\) inner product. In this representation, \(\psi(p_j)\) may be viewed as an effective wavefunction, with \(p_j\) playing the role of an effective position coordinate. Equation~\eqref{eigenmode} then becomes
\begin{equation}\label{eigenmode2}
-\frac{1}{2} f_{\mathrm{eq}}^{-1/2}\,
\frac{\partial}{\partial p^j}
\left[
f_{\mathrm{eq}}\,
\frac{\partial}{\partial p_j}
\left(f_{\mathrm{eq}}^{-1/2}\psi\right)
\right]
+\chi v^1 \psi
=E\psi\, .
\end{equation}
Finally, applying the product rule allows the left-hand side to be written in Schr\"odinger form,
\begin{equation}\label{eigenmode3}
\left[
-\frac{1}{2}\frac{\partial}{\partial p^j}\frac{\partial}{\partial p_j}
+ V_\chi(p_j)
\right]\psi
=E\psi\, ,
\end{equation}
where the effective potential is given by (compare with \cite[\S 4.9]{PavliotisBook})
\begin{equation}\label{effectivepotential}
\boxed{V_\chi(p_j)
=\dfrac{\beta^2 v^j v_j}{8}-\dfrac{\beta}{4} \dfrac{\partial v^j}{\partial p^j} +\chi v^1 \, .}
\end{equation}
We see that, in this formulation, the relaxation rate \(i\omega\) plays the role of the energy eigenvalue, while the complexified wavenumber $ik$ serves as a perturbative parameter. If the (quasi)particle dispersion relation \(\varepsilon=\varepsilon(p_j)\) is isotropic, the unperturbed potential \(V_0\) is likewise isotropic. In this case, the term proportional to \(ik\) acts as a perturbation that explicitly breaks rotational symmetry by selecting the \(x\)-direction.

We recall that $\Phi=\feq$ is a solution of the Boltzmann equation with $\omega=k=0$. Hence, as a consistency check, let us verify explicitly that $\psi=\sqrt{\feq}$ is a solution of \eqref{eigenmode3} with $E=\chi=0$:
\begin{equation}
\frac{1}{2}\frac{\partial}{\partial p^j}\frac{\partial}{\partial p_j}
\sqrt{\feq} =
\frac{\partial}{\partial p^j}\frac{\partial}{\partial p_j}\left(\frac{1}{2}
e^{\frac{\alpha}{2}-\frac{\beta\varepsilon}{2}}\right)=\frac{\partial}{\partial p^j}\left(-\dfrac{\beta v^j}{4}
e^{\frac{\alpha}{2}-\frac{\beta\varepsilon}{2}}\right)=\dfrac{\beta^2 v^jv_j}{8}
e^{\frac{\alpha}{2}-\frac{\beta\varepsilon}{2}}-\dfrac{\beta}{4} \dfrac{\partial v^j}{\partial p^j} e^{\frac{\alpha}{2}-\frac{\beta\varepsilon}{2}}=V_0 \sqrt{\feq} \, .  
\end{equation}

\subsection{Some notable effective potentials}\label{notablepotentials}
\vspace{-0.1cm}

Below, we collect the explicit form of the effective potential corresponding to free-particle dispersion relations.

\paragraph*{Nonrelativistic particles.}
For a Newtonian particle of mass \(m\) in \(D\) spatial dimensions, the single-particle dispersion relation is
\(\varepsilon=p^j p_j/(2m)\).
The corresponding effective potential is
\begin{equation}\label{Voscill}
V_\chi(p_j)
=
\frac{\beta^2\,p^j p_j}{8m^2}
-\frac{\beta D}{4m}
+\frac{\chi\,p^1}{m}\, .
\end{equation}
This potential describes a harmonic oscillator with frequency \(\beta/(2m)\), shifted by a constant so that the ground-state energy vanishes, in agreement with the Fokker-Planck literature \cite[\S 5.5.1]{RiskenFP}. The term proportional to \(\chi\) represents a uniform force \(\chi/m\) acting along the \(x^1\) direction.

\paragraph*{Relativistic particles.}
For a relativistic particle of mass \(m\) in \(D\) spatial dimensions, the dispersion relation is
\(\varepsilon=\sqrt{m^2+p^j p_j}\), and the effective potential takes the form
\begin{equation}\label{theUgly}
V_\chi(p_j)
=
\frac{\beta^2\, p^j p_j}{8\,(m^2+p^j p_j)}
-\frac{\beta}{4}\left[
\frac{D-1}{(m^2+p^j p_j)^{1/2}}
+\frac{m^2}{(m^2+p^j p_j)^{3/2}}
\right]
+\frac{\chi\, p^1}{(m^2+p^j p_j)^{1/2}} \, .
\end{equation}
Unlike the nonrelativistic case, this potential does not admit a simple mechanical interpretation.

\paragraph*{Massless particles.}
For massless particles in \(D\) spatial dimensions, \(\varepsilon=\sqrt{p^j p_j}\), and the effective potential is obtained by taking the ultrarelativistic limit \(m\to0\) of equation \eqref{theUgly}, yielding
\begin{equation}\label{ultrarelativisticpotential}
V_\chi(p_j)
=
\begin{cases}
\dfrac{\beta^2}{8}
-\dfrac{\beta}{2}\,\delta(p)
+\chi\,\mathrm{sgn}(p),
& D=1\, , \\[6pt]
\dfrac{\beta^2}{8}
-\dfrac{\beta(D-1)}{4\sqrt{p^j p_j}}
+\dfrac{\chi\, p^1}{\sqrt{p^j p_j}},
& D\ge 2\, .
\end{cases}
\end{equation}
In one spatial dimension, the effective potential consists of an attractive Dirac delta function (compare with \cite[\S 5.5.4]{RiskenFP}) supplemented by an odd step-function perturbation. In contrast, for \(D\ge2\) the potential exhibits a Coulomb-like \(1/r\) behavior (compare with \cite{Felderhof2012}), perturbed by a dipolar term proportional to \(x^1/r\).

The qualitative difference between the one- and higher-dimensional cases originates from the behavior of the term proportional to \(m^2\) in  \eqref{theUgly}. Indeed, in one dimension,
\begin{equation}
\int_{-P}^{P}\frac{m^2\,dp}{(m^2+p^2)^{3/2}}
=\frac{2P}{\sqrt{m^2+P^2}}
\xrightarrow{m\to0} 2\, ,
\end{equation}
so that this contribution converges to \(2\delta(p)\), whereas in \(D\ge2\) the corresponding integral vanishes as $m\rightarrow 0$, and no Dirac delta term is generated. We remark that the same result may alternatively be obtained by evaluating equation \eqref{effectivepotential} directly for the massless dispersion relation \(\varepsilon=\sqrt{p^j p_j}\). In this case, the velocity field \(v^j\) is smooth for \(p^j\neq0\), so that
\(
\partial v^j/\partial p^j=(D-1)/\varepsilon
\)
away from the origin. At \(p^j=0\), however, the derivative becomes ill-defined and must be understood in the sense of distributions. Applying the divergence theorem, one finds
\begin{equation}
\int_{\mathcal{S}_D(P)} \frac{\partial v^j}{\partial p^j}\, d^D p
=
\int_{\partial\mathcal{S}_D(P)} v^j\, d^{D-1}S_j
=
\int_{\partial\mathcal{S}_D(P)} d^{D-1}S
=
\mathrm{Area}\!\left(\partial\mathcal{S}_D(P)\right),
\end{equation}
with $\mathcal{S}_D(P)$ the $D$-sphere of radius $P$.
In the limit \(P{\to}\,0\), the surface area converges to $2$ for $D\,{=}\,1$ and to $0$ for $D\,{\geq}\,2$.

\vspace{-0.1cm}
\subsection{Quasi-normal modes of Newtonian Fokker-Planck kinetic theory}\label{NewtonianCase}
\vspace{-0.1cm}

\begin{figure}[h!]
    \centering
\includegraphics[width=0.4\linewidth]{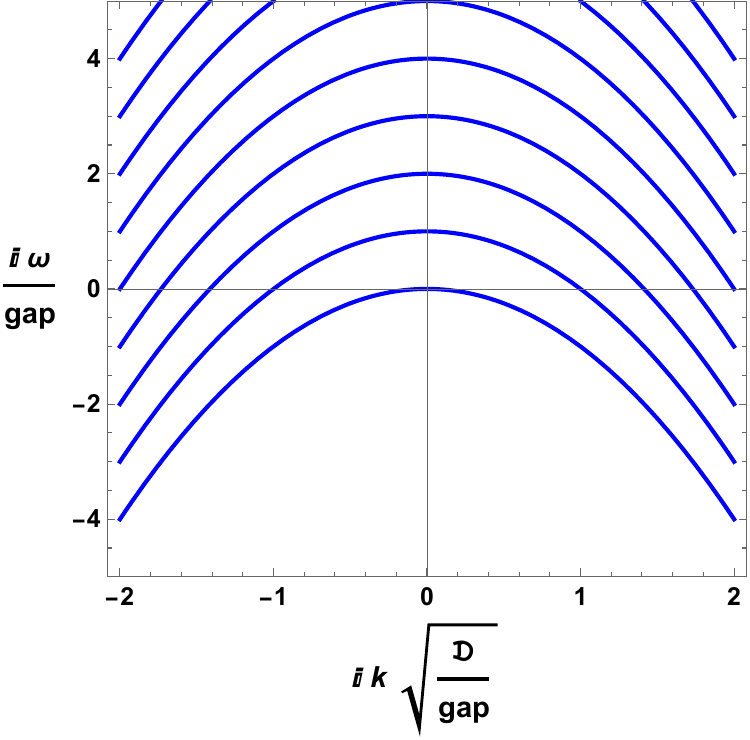}\hspace{0.08\linewidth}
\includegraphics[width=0.4\linewidth]{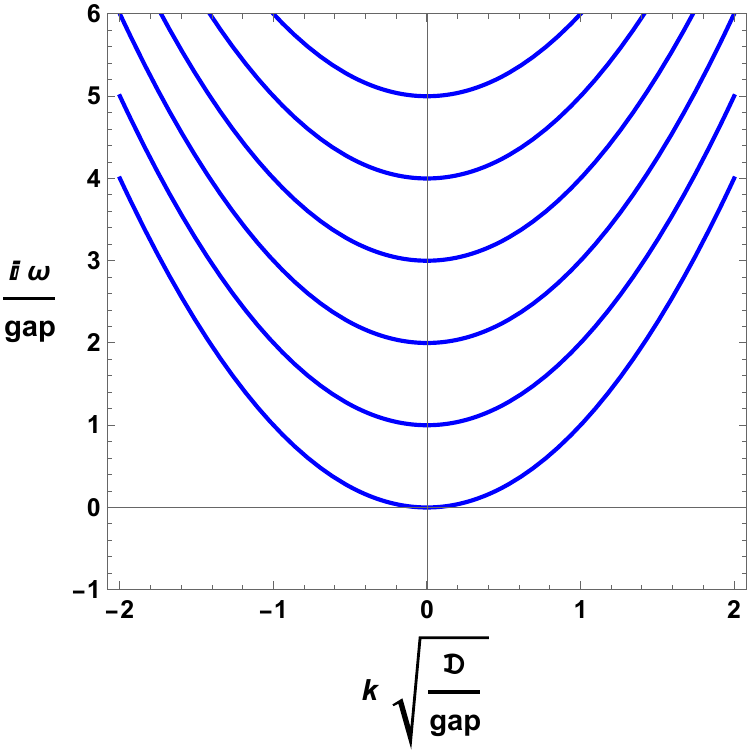}
\caption{Quasi-normal frequencies of nonrelativistic Fokker-Planck kinetic theory for imaginary \(k\) (left) and real \(k\) (right), in units of the spectral gap $\beta\nu/m$. The hydrodynamic mode is unique and nondegenerate, and exhibits an exact diffusive dispersion relation \(\omega=-i\mathfrak{D}k^2\) at all \(k\), with diffusion coefficient \(\mathfrak{D}=(\beta^2\nu)^{-1}\). In addition, an infinite tower of nonhydrodynamic modes is present, equally spaced by \(\nu\beta/m\) and degenerate for \(D\ge2\). For real \(k\), all modes remain purely damped, i.e.\ \(\omega\in i\mathbb{R}\), in sharp contrast with RTA \cite{RomatschkeCutsandPoles:2015gic,Bajec:2024jez,Brants:2024wrx}, where the nonhydrodynamic modes form a continuous line \(\omega\in\mathbb{R}-i\times\mathrm{gap}\). This qualitative difference originates from the microscopic dynamics encoded in the collision operator: in RTA, particles propagate ballistically between collisions, whereas the Fokker-Planck operator describes diffusion in momentum space induced by many small-angle scatterings, which in turn enforces diffusive motion in real space at all scales (at least in the Newtonian case).
}
    \label{fig:Newtonian}
\end{figure}

Introducing the unit vector \(n^j=(1,0,0,\ldots)\), the potential in Eq.~\eqref{Voscill} can be rewritten in the equivalent form
\begin{equation}\label{Voscill2}
V_\chi(p_j)
=
\frac{\beta^2}{8m^2}
\left(p^j\,{+}\,\frac{4m\chi}{\beta^2}n^j\right)
\left(p_j\,{+}\,\frac{4m\chi}{\beta^2}n_j\right)
-\frac{\beta D}{4m}
-\frac{2\chi^2}{\beta^2}\, .
\end{equation}
This expression describes a \(D\)-dimensional harmonic oscillator in momentum space, translated along the \(-x^1\) direction by an amount \(4m\chi/\beta^2\) and shifted downward by a constant. The corresponding quasi-normal spectrum is therefore discrete and given by
\begin{equation}
E_N=\frac{\beta}{2m}\,N-\frac{2\chi^2}{\beta^2}
\qquad\Longrightarrow\qquad
i\omega_N=\frac{\nu\beta}{m}\,N+\frac{k^2}{\beta^2\nu},
\qquad N\in\mathbb{N}\, .
\end{equation}
Strictly speaking, the correspondence with a harmonic oscillator holds for real \(\chi\) (i.e.\ imaginary \(k\)). Nevertheless, upon complexifying \(\chi\), the Hermite eigenfunctions remain normalizable and continue to satisfy \eqref{eigenmode3} by analytic continuation. The dispersion relations above therefore extend to arbitrary complex \(k\), and are plotted in figure \ref{fig:Newtonian}.

\section{Ultrarelativistic particles in one dimension}

We are finally entering the central part of the paper, where we determine the quasi-normal modes of ultrarelativistic Fokker-Planck kinetic theory, and compare the resulting spectrum with the Newtonian case (shown in figure \ref{fig:Newtonian}). The mathematical analysis is more involved than in the Newtonian case (discussed in section \ref{NewtonianCase}), but the problem ultimately boils down to computing the spectra of non-Hermitian Hamiltonians. In this section, we focus on the one-dimensional problem, corresponding to the case \(D=1\) in \eqref{ultrarelativisticpotential}.

\subsection{Setup of the problem}

We seek solutions \(\psi(p)\) of the differential equation
\begin{equation}\label{Schrod1D}
-\frac{1}{2}\psi''
=
\left[
E-\frac{\beta^2}{8}
+\frac{\beta}{2}\,\delta(p)
-\chi\,\mathrm{sgn}(p)
\right]\psi
\end{equation}
that remain bounded as \(|p|\to\infty\) (if $\psi$ is square integrable, the corresponding $E$ belongs to the point spectrum; if $\psi$ is oscillatory, $E$ belongs to the continuous spectrum \cite[\S 21]{landau3}). This defines a standard one-dimensional Schr\"odinger problem, with the important difference that \(\chi\) may in general be complex, rendering the Hamiltonian non-Hermitian. Nevertheless, the construction of solutions proceeds along familiar lines. In particular, equation \eqref{Schrod1D} can be treated piecewise as
\begin{equation}\label{positiveandnegative}
\begin{aligned}
-\frac{1}{2}\psi''
&=
\left(E-\frac{\beta^2}{8}+\chi\right)\psi,
&& p<0, \\
-\frac{1}{2}\psi''
&=
\left(E-\frac{\beta^2}{8}-\chi\right)\psi,
&& p>0,
\end{aligned}
\end{equation}
supplemented by the matching conditions at \(p=0\),
\begin{equation}\label{matchingconditions}
\psi(0^+)=\psi(0^-),\qquad
\psi'(0^+)-\psi'(0^-)=-\beta\,\psi(0),
\end{equation}
together with the requirement that \(\psi\) remain finite as \(p\to\pm\infty\).

\subsection{General form of the solutions}
\vspace{-0.1cm}

The general solution of equation \eqref{positiveandnegative} can be written as
\begin{equation}\label{splittuzzionenne}
\psi(p)=
\begin{cases}
\mathcal{A}\,e^{-(r_1+i s_1)p}+\mathcal{B}\,e^{(r_1+i s_1)p}, & p<0, \\[4pt]
\mathcal{C}\,e^{-(r_2+i s_2)p}+\mathcal{D}\,e^{(r_2+i s_2)p}, & p>0,
\end{cases}
\end{equation}
where \(\mathcal{A},\mathcal{B},\mathcal{C},\mathcal{D}\in\mathbb{C}\) are arbitrary constants and
\(r_1,s_1,r_2,s_2\in\mathbb{R}\) are related to $E$ and $\chi$ via
\begin{equation}\label{EnadChiin1D}
\begin{aligned}
E
&=\frac{\beta^2}{8}
-\frac{1}{4}\!\left[
r_1^2+r_2^2-s_1^2-s_2^2
+2i\left(r_1 s_1+r_2 s_2\right)
\right], \\
\chi
&=-\frac{1}{4}\!\left[
r_1^2-r_2^2-s_1^2+s_2^2
+2i\left(r_1 s_1-r_2 s_2\right)
\right].
\end{aligned}
\end{equation}
By varying \(r_1,s_1,r_2,s_2\) and imposing the boundary and matching conditions through appropriate choices of \(\mathcal{A},\mathcal{B},\mathcal{C},\mathcal{D}\), the full quasi-normal-mode spectrum can be explored. Without loss of generality, we will take \(r_1,r_2\ge0\).

\vspace{-0.1cm}
\subsection{The hydrodynamic mode}
\vspace{-0.1cm}

To determine the discrete part of the spectrum, we look for physical eigenfunctions, i.e. for states belonging to the Hilbert space. In the present context, this is equivalent to requiring that $\psi$ decay at spatial infinity. This condition enforces $r_1,r_2>0$, which in turn implies $\mathcal{A}=\mathcal{D}=0$. Continuity of the wavefunction at the origin then requires $\mathcal{B}=\mathcal{C}$. By linearity, we may fix their common value to unity, yielding the bound-state ansatz
\begin{equation}\label{hydroModduzzoin1D}
\psi(p)=
\begin{cases}
e^{(r_1+i s_1)p}, & p<0, \\[4pt]
e^{-(r_2+i s_2)p}, & p>0 .
\end{cases}
\end{equation}

It remains to impose the second matching condition in \eqref{matchingconditions}, which leads to the constraints $r_1+r_2=\beta$ and $s_1+s_2=0$. Substituting these relations into \eqref{EnadChiin1D}, we obtain
\begin{equation}\label{EnadChiin1DBound}
E=-\frac{1}{2}\!\left( r_1
+ i s_1-\dfrac{\beta}{2}
\right)^2, \qquad \qquad
\chi
=-\frac{\beta}{2}\!\left( r_1
+ i s_1-\dfrac{\beta}{2}
\right).
\end{equation}
Combining these expressions yields a single hydrodynamic dispersion relation, valid for complex $k$,
\begin{equation}\label{diffondiamoinunadimensioneullalla}
\omega=-i\,\dfrac{k^2}{\beta^2\nu}\, .
\end{equation}
This is precisely the same diffusive mode encountered in the Newtonian theory, with an identical diffusion coefficient. There is, however, an important subtlety: this mode does not exist for arbitrary complex $k$. Indeed, since $r_1,r_2>0$ and $r_1+r_2=\beta$, equation \eqref{EnadChiin1DBound} implies the bound
\begin{equation}\label{consistenziamo}
-\dfrac{\beta^2\nu}{2}
< \mathfrak{Im}\, k
< \dfrac{\beta^2\nu}{2} \, .
\end{equation}

In summary, the discrete spectrum consists of a single dispersion relation (the hydrodynamic mode) which is well defined for all real values of $k$, but ceases to exist once $|\mathfrak{Im}\, k|< \beta^2\nu/2$ is violated.

\vspace{-0.1cm}
\subsection{Purely oscillatory wavefunctions}\label{oscillamus}
\vspace{-0.1cm}

All nonhydrodynamic modes belong to the continuous spectrum. In the present setting, this implies that the associated wavefunctions do not decay as \(p\to\pm\infty\), but instead remain oscillatory in at least one direction. Let us focus on solutions that oscillate for both \(p<0\) and \(p>0\), corresponding \(r_1=r_2=0\). In this case, \eqref{EnadChiin1D} reduces to
\begin{equation}\label{EnadChiin1DOscill}
E=\frac{\beta^2}{8}+\frac{s_1^2+s_2^2}{4},
\qquad
\chi=\frac{s_1^2-s_2^2}{4}.
\end{equation}
Inspection of the matching conditions in equation \eqref{splittuzzionenne} shows that, for any \((s_1,s_2)\in\mathbb{R}^2\), there exists a choice of coefficients \(\mathcal{A},\mathcal{B},\mathcal{C},\mathcal{D}\) satisfying all constraints. Consequently, every pair \((E,\chi)\in\mathbb{R}^2\) with \(E\ge \beta^2/8+|\chi|\) corresponds to a nonhydrodynamic mode. This implies the presence of a continuous band of quasi-normal modes with \(k\in i\mathbb{R}\) and
\begin{equation}
i\omega \ge \frac{\beta^2\nu}{4}+|k| \, .
\end{equation}

\subsection{Hybrid cases}
The remaining case corresponds to hybrid solutions in which the wavefunction decays on one half-line and remains oscillatory on the other, i.e.\ either \(r_1>0\) and \(r_2=0\) or \(r_1=0\) and \(r_2>0\). We analyze explicitly the first situation, \(r_2=0\), as the second is entirely analogous.

Substituting \(r_2=0\) into \eqref{EnadChiin1D}, we obtain
\begin{equation}\label{EnadChiin1DHybrid}
\begin{aligned}
E
&=\frac{\beta^2}{8}
-\frac{1}{4}\!\left[(r_1+i s_1)^2-s_2^2\right], \\
\chi
&=-\frac{1}{4}\!\left[(r_1+i s_1)^2+s_2^2\right],
\end{aligned}
\end{equation}
which implies the relation \(E=\beta^2/8+\chi+s_2^2/2\). The problem, therefore, reduces to determining the allowed values of \(\chi\) and \(s_2\).
Imposing continuity of the wavefunction at the origin yields
\begin{equation}\label{assssymmmetriccc_cases}
\psi(p)=
\begin{cases}
\mathcal{B}\,e^{(r_1+i s_1)p}, & p<0, \\[4pt]
\mathcal{B}\,\cos(s_2 p)+\mathcal{F}\,\sin(s_2 p), & p>0,
\end{cases}
\end{equation}
with \(\mathcal{F}\in\mathbb{C}\). The derivative matching condition in \eqref{matchingconditions} then gives
\(
s_2\mathcal{F}-(r_1+i s_1)\mathcal{B}=-\beta\mathcal{B}.
\)
If \(\mathcal{B}=0\), this relation forces \(s_2\mathcal{F}=0\), leading to the trivial solution \(\psi\equiv0\). We may therefore assume \(\mathcal{B}\neq0\) and, by linearity, set \(\mathcal{B}=1\), leaving
\(
s_2\mathcal{F}=r_1-\beta+i s_1.
\)
This condition can be satisfied for arbitrary \(r_1>0\), \(s_1\in\mathbb{R}\), and \(s_2\neq0\). Since the spectrum of the operator is closed,\footnote{A self-adjoint operator perturbed by a bounded operator is closed, and closed operators have closed spectrum.} the limiting case \(s_2=0\) is also included by continuity. As a result, one obtains the quasi-normal band
\(
E\in \chi+[\beta^2/8,+\infty)
\),
or equivalently
\begin{equation}\label{labandinanna}
i\omega \in ik+\bigg[\frac{\beta^2\nu}{4},+\infty\bigg).
\end{equation}

Finally, we address the range of admissible \(\chi\). Since \(r_1+i s_1\) may be any complex number with positive real part, \((r_1+i s_1)^2\) spans \(\mathbb{C}\setminus(-\infty,0]\). The present construction therefore does not cover the case \(\chi>0\); however, this regime was already analyzed in section \ref{oscillamus} and yields the same spectral band \eqref{labandinanna}. Repeating the analysis for the complementary case \(r_1=0\) and \(r_2>0\) leads to the symmetric band
\begin{equation}\label{labandinanna2}
i\omega \in -ik+\bigg[\frac{\beta^2\nu}{4},+\infty\bigg).
\end{equation}

\subsection{Putting everything together}

\begin{figure}[h!]
    \centering
\includegraphics[width=0.4\linewidth]{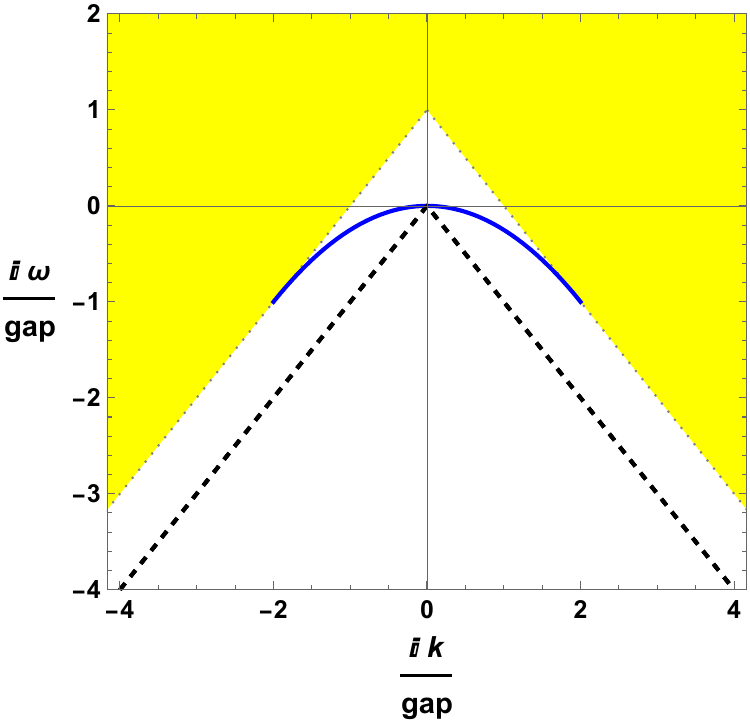}\hspace{0.08\linewidth}
\includegraphics[width=0.4\linewidth]{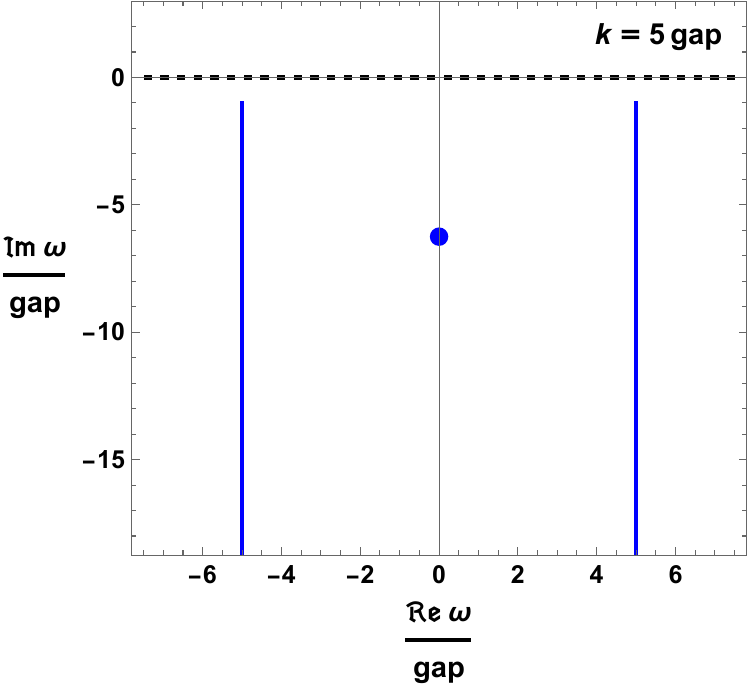}
\caption{Quasi-normal frequencies of ultrarelativistic Fokker-Planck kinetic theory in one spatial dimension, for imaginary \(k\) (left) and real \(k\) (right), all in units of \(\beta^2\nu/4\) (the spectral gap at $k{=}0$). Left panel: A single hydrodynamic mode exhibits diffusive dispersion, \(\omega{=}-i\mathfrak{D}k^2\), with diffusion coefficient \(\mathfrak{D}=(\beta^2\nu)^{-1}\). All remaining modes form a continuum (yellow region). The diffusive mode merges into the continuum before being able to cross the line \(\mathfrak{Im}\omega\,{=}\,|\mathfrak{Im}k|\) (dashed), whose violation would signal acausality \cite{HellerBounds2022ejw,GavassinoBounds2023myj}. Right panel: For real \(k\), the hydrodynamic mode is imaginary, while the nonhydrodynamic modes organize into two vertical continua emerging from \((\pm k,-i\,\text{gap})\). The appearance of propagating nonhydrodynamic modes in the relativistic theory, in contrast with the purely damped Newtonian spectrum, reflects the underlying microscopic dynamics: momentum diffusion in the Newtonian case leads to overdamped Brownian motion, whereas in the ultrarelativistic one-dimensional case the discrete velocity spectrum (\(v{=}\pm1\)) permits ballistic propagation despite stochastic momentum exchange.
}
    \label{fig:1D}
\end{figure}

Having analyzed all admissible solution sectors, we have thus determined the complete quasi-normal-mode spectrum of the theory. The resulting spectra are displayed in figure \ref{fig:1D}, respectively for imaginary wavenumber \(k\) (left panel) and for real \(k\) (right panel). For purely imaginary wavenumber, \(k\in i\mathbb{R}\), the spectrum can be written explicitly as
\begin{equation}\label{1dImaginaryKexactformulaiomega}
i\omega\in
\begin{cases}
\displaystyle
\left\{-\frac{(ik)^2}{\beta^2\nu}\right\}
\;\cup\;
\bigg[\dfrac{\beta^2\nu}{4}-|ik|,\,+\infty\bigg),
& |ik|<\dfrac{\beta^2\nu}{2}, \\[10pt]
\displaystyle
\bigg[\dfrac{\beta^2\nu}{4}-|ik|,\,+\infty\bigg),
& |ik|\ge\dfrac{\beta^2\nu}{2}.
\end{cases}
\end{equation}
In this case, the hydrodynamic diffusive mode \(\omega=-i k^2/(\beta^2\nu)\) is present only for sufficiently small \(|k|\), while for larger imaginary wavenumber it merges into the nonhydrodynamic continuum. The remaining modes form a continuous band bounded from below, corresponding to oscillatory or hybrid solutions of the effective Schr\"odinger problem. For real wavenumber, \(k\in\mathbb{R}\), the quasi-normal spectrum instead takes the form
\begin{equation}
\omega \in
\left\{-\,i\frac{k^2}{\beta^2\nu}\right\}
\;\cup\;
k-i\bigg[\frac{\beta^2\nu}{4},+\infty\bigg)
\;\cup\;
-k-i\bigg[\frac{\beta^2\nu}{4},+\infty\bigg).
\end{equation}
In this case, the hydrodynamic mode remains purely diffusive for all \(k\), while the nonhydrodynamic modes organize into two continuous branches with real parts \(\pm k\) and uniform damping bounded from below by \(\beta^2\nu/4\). These branches correspond to propagating, damped excitations, reflecting the persistence of inertial transport in the relativistic Fokker-Planck dynamics.

\section{Ultrarelativistic particles in three dimensions}

In this final section, we analyze the quasi-normal-mode spectrum of ultrarelativistic Fokker-Planck kinetic theory in three spatial dimensions, corresponding to \(D=3\) in equation \eqref{ultrarelativisticpotential}. While a complete analytic determination of the spectrum at finite \(k\) is not available in this case, the hydrodynamic mode and the continuous spectrum can still be obtained in closed form, and the remainder of the spectrum can be constrained using standard tools from functional analysis.

\subsection{Spectrum at zero wavenumber}

We look for solutions $\psi(p_1,p_2,p_3)$ of the Schr\"odinger equation
\begin{equation}\label{schrod3D}
\left[
-\frac{1}{2}\frac{\partial}{\partial p^j}\frac{\partial}{\partial p_j}
+\frac{\beta^2}{8}
-\frac{\beta}{2p}
+\frac{\chi\, p^1}{p}
\right]\psi
=E\psi
\end{equation}
that remain bounded in the limit $p\equiv\sqrt{p^j p_j}\to\infty$. For $\chi=0$, the problem reduces to that of the hydrogen atom, with the Coulomb potential shifted upward by a constant $\beta^2/8$. It follows that, at vanishing wavenumber, the spectrum consists of an infinite list of discrete points, together with a continuous branch, given by
\begin{equation}\label{hydraUU}
\begin{split}
E_N ={}& \frac{\beta^2}{8}\left(1-\frac{1}{N^2}\right)
\qquad \Longrightarrow \qquad
i\omega_N = \frac{\beta^2\nu}{4}\left(1-\frac{1}{N^2}\right),
\qquad N\in\mathbb{N},\\
E \in{}& \bigg[\frac{\beta^2}{8},+\infty\bigg)
\qquad\qquad\;\;\;
\Longrightarrow \qquad
i\omega \in \bigg[\frac{\beta^2\nu}{4},+\infty\bigg).
\end{split}
\end{equation}
In particular, there is a unique, nondegenerate hydrodynamic mode with \(i\omega_1=0\). The slowest nonhydrodynamic modes arise from the \(N=2\) Coulomb level, which is fourfold degenerate (reflecting the \(N^2\) hydrogenic degeneracy) and has relaxation rate \(i\omega_2=3\beta^2\nu/16\). At finite \(k\), some of this degeneracy is expected to be lifted, since the term \(p^1/p\) explicitly breaks isotropy, leaving the same residual degeneracy as in the Stark effect \cite[\S 76]{landau3}.

Our goal is to determine how the spectrum is modified upon introducing the dipolar perturbation \(\chi\, p^1/p\), allowing for complex values of \(\chi\).

\subsection{The hydrodynamic mode}
\vspace{-0.2cm}

One may verify directly that the ansatz
\vspace{-0.1cm}
\begin{equation}
\psi = e^{-\frac{\beta p}{2}-\frac{2\chi p^1}{\beta}}
\end{equation}
is an exact eigenfunction of \eqref{schrod3D}, with eigenvalue \(E=-2\chi^2/\beta^2\). This solution yields the diffusive dispersion relation \(\omega=-ik^2/(\beta^2\nu)\). As can be seen, the diffusivity \(\mathfrak{D}=(\beta^2\nu)^{-1}\) coincides with that obtained in the one-dimensional analysis, cf. \eqref{diffondiamoinunadimensioneullalla}, which in turn agrees with the Newtonian result. As before, imposing boundedness of \(\psi\) at infinity leads to the constraint \(|\mathfrak{Re}\,\chi|\leq \beta^2/4\), or equivalently \(|\mathfrak{Im}\,k|\leq \beta^2\nu/2\), which is precisely the same consistency condition \eqref{consistenziamo} encountered in the one-dimensional case.

\vspace{-0.3cm}
\subsection{The continuous part of the spectrum}
\vspace{-0.3cm}

For Schr\"odinger-type operators with bounded potentials that depend smoothly on the angular variables, the continuous spectrum can be characterized using a standard argument due to Weyl \cite[Lemma 6.17]{TeschlBook}. One considers a compactly supported wavefunction $\psi$, and translates it arbitrarily far in a fixed direction. As the support of $\psi$ is pushed toward spatial infinity, the potential approaches a direction-dependent constant, so that $\psi$ asymptotically behaves as a free particle subject to this effective potential. The corresponding free-particle energies generate a subset of the continuous spectrum. Repeating this construction for all directions reconstructs the full continuous spectrum.

In the present case, a wavefunction localized at infinity experiences the asymptotic potential
\(
V_\chi=\beta^2/8+\chi\cos\theta,
\)
where $\theta$ denotes the angle between the $p^1$ axis and the direction along which the wavefunction is translated. The associated spectral contribution is therefore
\(
E\in \chi\cos\theta+[\beta^2/8,+\infty),
\)
a result that remains valid even for complex $\chi$. Taking the union over all angles and mapping back to frequencies and wavevectors yields
\vspace{-0.1cm}
\begin{equation}
\omega\in k\, [-1,1]-i\bigg[\frac{\beta^2\nu}{4},+\infty\bigg)\, .
\end{equation}
For any fixed $k\notin i\mathbb{R}$, this describes a two-dimensional region in the complex $\omega$ plane whose extent in $\mathfrak{Re}\,\omega$ coincides with that of the relativistic relaxation-time approximation, while extending to $-\infty$ along the $\mathfrak{Im}\,\omega$ direction. 

Note that, when \(k\in i\mathbb{R}\), the combined contribution of the hydrodynamic mode and the continuous branch reproduces exactly the spectrum of the one-dimensional system given in equation \eqref{1dImaginaryKexactformulaiomega}, see figures \ref{fig:1D} and \ref{fig:3D}.

\vspace{-0.3cm}
\subsection{Analytical constraints on the discrete non-hydrodynamic modes}
\vspace{-0.3cm}

The discrete nonhydrodynamic eigenvalues at finite wavenumber are not as easily accessible. Nevertheless, since the perturbation $\chi\,p^1/p$ is a bounded operator with operator norm $|\chi|$, several general constraints on the mode structure can be established. For example, all discrete eigenvalues are guaranteed to be analytic in a neighborhood of $\chi=0$ \cite{Kato1949_Perturbation_I}. They therefore admit convergent power-series expansions in $\chi$, and standard perturbative methods apply, although we will not pursue this analysis explicitly here.

A second, and more informative, constraint follows from general results on bounded perturbations of self-adjoint operators. When a self-adjoint operator is perturbed by a bounded (not necessarily self-adjoint) operator of norm $|\chi|$, the spectrum of the perturbed operator must lie within a distance $|\chi|$ in the complex plane from the unperturbed spectrum \cite[\S 5.4.3]{Kato_Perturbation_Theory}. In the present case, the unperturbed spectrum consists of the set $\bigcup_N \{E_N\} \cup [\beta^2/8,+\infty)$ given in equation \eqref{hydraUU}. It follows that the discrete eigenvalues at finite $\chi$ must lie within the closed $|\chi|$-neighborhood of this set in the complex plane. The region excluded by this criterion (and by stability) is marked in red in figure \ref{fig:3D}.

\vspace{-0.3cm}
\subsection{About ballistic propagation}\label{WhyBallistic}
\vspace{-0.2cm}

We recall that, in the Newtonian case (figure \ref{fig:Newtonian}), the spectrum for real $k$ is purely damped, $\omega\in i\mathbb{R}$, in contrast with the relaxation-time approximation, where a continuous ballistic branch is present. This difference originates from the underlying microscopic dynamics. In the Newtonian Fokker-Planck theory, random momentum kicks $\Delta p^j$ induce velocity changes $\Delta v^j=\Delta p^j/m$, leading to diffusive spreading of perturbations and suppressing ballistic propagation.
In the one-dimensional ultrarelativistic case, by contrast, the velocity takes only the discrete values $v=\pm1$. Small momentum transfers therefore do not generically change the particle velocity, and ballistic motion is preserved, giving rise to two continuous spectral branches with $\mathfrak{Re}\,\omega=\pm k$ (figure \ref{fig:1D}).

\begin{figure}[h!]
    \centering
\includegraphics[width=0.4\linewidth]{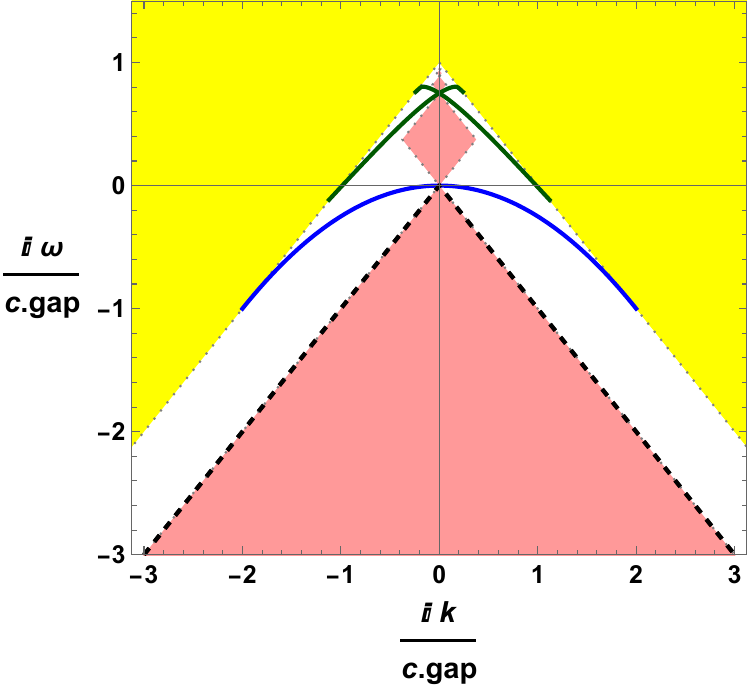}\hspace{0.08\linewidth}
\includegraphics[width=0.4\linewidth]{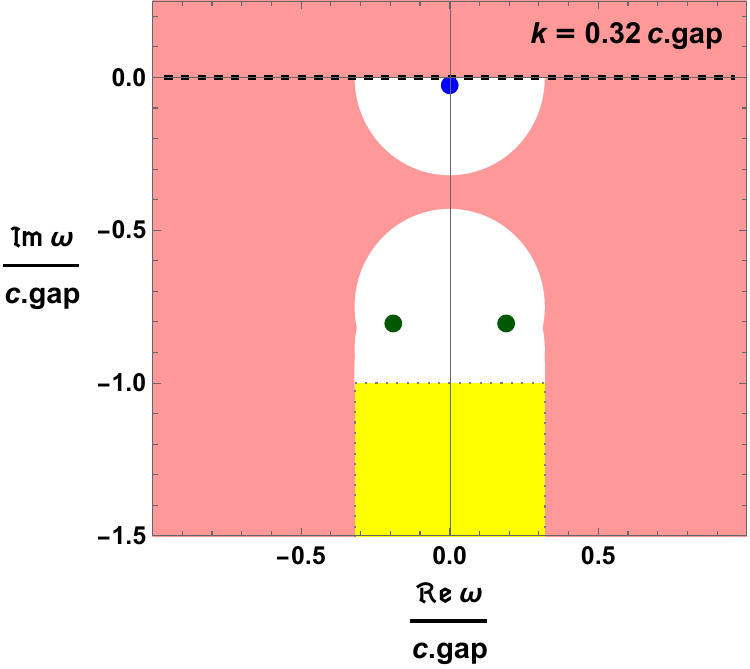}
\caption{Quasi-normal frequencies of ultrarelativistic Fokker-Planck kinetic theory in three spatial dimensions, for imaginary \(k\) (left) and real \(k\) (right), all expressed in units of $\beta^2\nu/4$ (the spectral gap of the continuous band, ``c.gap'', at $k=0$). Left panel: A single hydrodynamic mode exhibits diffusive dispersion, \(\omega=-i\mathfrak{D}k^2\), with diffusion coefficient \(\mathfrak{D}=(\beta^2\nu)^{-1}\). Above it, we find an infinite tower of discrete modes that we could not determine analytically, but two of which (specifically, two of the three modes emerging from the splitting of the first non-hydrodynamic level $\omega_2$) have been calculated numerically for illustration (green). At $k=0$, the spectral degeneracy is the same as that of the hydrogen atom, and just like the hydrogen atom, these levels undergo Stark splitting at finite $k$. The remaining modes form a continuum (yellow region). Again, the diffusive mode merges into the continuum before it gets a chance of crossing the boundary of stability \(\mathfrak{Im}\omega=|\mathfrak{Im}k|\) (dashed). Right panel: For real \(k\), the hydrodynamic mode is imaginary, while the discrete non-hydrodynamic modes may acquire a real part, as shown by the two numerical examples (green). The continuous branch forms an infinitely long vertical rectangle with upper corners \((\pm k,-i\,\text{c.gap})\) (yellow), signalling the presence of ballistic propagation, whose existence is discussed in section \ref{WhyBallistic}. In both panels, the shade of red marks the forbidden region, where we are guaranteed \textit{not to find} other modes.}
    \label{fig:3D}
\end{figure}

A similar mechanism persists in three-dimensional space. The continuous spectrum is controlled by the large-momentum sector. In this limit, momentum fluctuations induce velocity changes $\Delta v^k=(\delta^k_j-v^k v_j)\Delta p^j/\varepsilon\to0$, so that the velocity becomes asymptotically conserved. Relativistic Fokker-Planck kinetics therefore admits ballistic transport, resulting in a continuous band of modes with $\mathfrak{Re}\,\omega\in[-k,k]$ (figure \ref{fig:3D}).

\newpage

\section{Conclusions}

In this work, we have determined the quasi-normal-mode spectrum of ultrarelativistic kinetic theories with a Fokker-Planck collision integral \cite{Debbasch,DunkelHanggi}. By exploiting the unitary equivalence between detailed-balanced Fokker-Planck operators and Schr\"odinger-type Hamiltonians, we recast the kinetic eigenvalue problem at fixed wavenumber as a quantum spectral problem with a perturbation proportional to the wavenumber. This correspondence provides a unified framework for analyzing both hydrodynamic and nonhydrodynamic excitations.

A central result of our analysis is that the hydrodynamic sector is governed by a single diffusive mode with dispersion relation $\omega=-i k^{2}/(\beta^{2}\nu)$, identical in one and three spatial dimensions. This mode exists only within a finite strip in the complex wavenumber plane, $|\mathfrak{Im}\,k|\le\beta^{2}\nu/2$, beyond which it merges into the nonhydrodynamic continuum. These explicit bounds enforce covariant stability and ensure that the diffusive mode never violates the relativistic causality constraint $\mathfrak{Im}\,\omega\leq|\mathfrak{Im}\,k|$ \cite{HellerBounds2022ejw,GavassinoBounds2023myj}.

In one spatial dimension, the remainder of the spectrum consists entirely of continuous branches. For real wavenumbers, these organize into two vertical continua with $\mathfrak{Re}\,\omega=\pm k$, signalling ballistic propagation despite stochastic momentum exchange. In three spatial dimensions, the unperturbed problem reduces to the hydrogen atom, yielding an infinite tower of discrete nonhydrodynamic modes together with a continuous band. At finite wavenumber, we characterized the full continuous spectrum using Weyl-type arguments and showed that relativistic Fokker-Planck kinetics generically admits a ballistic continuum with $\mathfrak{Re}\,\omega\in[-k,k]$.

Our results clarify how relativistic kinematics qualitatively modifies transient dynamics in kinetic theories with momentum-space diffusion. While stochastic scattering is present at the microscopic level, relativistic velocity saturation suppresses velocity randomization at high energy, allowing ballistic propagation to survive in the nonhydrodynamic sector. At the same time, the hydrodynamic response remains strictly diffusive, with stability bounds on its domain of validity.

Finally, these spectral results play a crucial role in a companion work \cite{GavassinoDiffusionCompatible:2026tvy}, where we use them to construct explicit solutions of relativistic Fokker-Planck kinetic theory whose associated particle density obeys the diffusion equation exactly, even at finite wavelength. Together, these works provide a coherent and fully relativistic account of how diffusion and ballistic transport coexist in stochastic kinetic systems.

\section*{Acknowledgements}

This work is supported by a MERAC Foundation prize grant,  an Isaac Newton Trust Grant, and funding from the Cambridge Centre for Theoretical Cosmology.

\appendix

\section{Derivation of the properties of the collision operator}\label{aaa}

In kinetic theory, the current densities of (quasi)particles, energy, and entropy, are given by\footnote{In relativistic thermodynamics, the energy current density is defined as $W^\mu{=}-\mathcal{K}_\nu T^{\mu \nu}$, where $\mathcal{K}^\nu$ is the timelike Killing vector that defines the velocity of the environment \cite{GavassinoGibbs2021}. In our case, the environment is the bath, which is at rest in Minkowski space, so $\mathcal{K}^\nu=(1,0,0,...)$.}
\begin{equation}
J^\mu =\int \dfrac{d^D p}{(2\pi)^D}  
\begin{bmatrix}
1 \\
v^j \\
\end{bmatrix} f \, ,
\qquad
W^\mu =\int \dfrac{d^D p}{(2\pi)^D}  
\begin{bmatrix}
1 \\
v^j \\
\end{bmatrix} f\varepsilon \, ,
\qquad
s^\mu =\int \dfrac{d^D p}{(2\pi)^D}  
\begin{bmatrix}
1 \\
v^j \\
\end{bmatrix} f(1-\ln f) \, .
\end{equation}
Now, if the gas were an isolated system, the energy-momentum flux \(W^\mu\) would be conserved, and the entropy current \(s^\mu\) would have non-negative divergence. In the present setting, however, the surrounding medium acts as a thermal bath, allowing for the exchange of energy and entropy with the environment. Thermodynamics nevertheless implies that the free energy of a system coupled to a heat bath is non-increasing \cite[\S 20]{landau5}, which can be expressed as
\(
\partial_\mu (s^\mu-\beta W^\mu)\ge 0 ,
\)
where \(\beta\) denotes the inverse temperature of the bath. In addition, the (quasi)particle current \(J^\mu\) is conserved, implying
\(
\partial_\mu z^\mu \ge 0 ,
\)
where the Massieu current is defined as
\(
z^\mu=s^\mu+\alpha J^\mu-\beta W^\mu .
\)
Writing this current explicitly and decomposing the distribution function as \(f=f_{\mathrm{eq}}+\Phi\), one finds \cite{GavassinoCausality2021,RochaGavassinoFlucut:2024afv}
\begin{equation}
z^\mu =\int \dfrac{d^D p}{(2\pi)^D}  
\begin{bmatrix}
1 \\
v^j \\
\end{bmatrix} f(1-\ln f+\ln \feq)=\int \dfrac{d^D p}{(2\pi)^D \feq}  
\begin{bmatrix}
1 \\
v^j \\
\end{bmatrix} \left(\feq^2-\tfrac{1}{2}\Phi^2\right) +\mathcal{O}(\Phi^3) \, .
\end{equation}
Taking the divergence of this equation and using \eqref{BoltzmannEquation}, one finds
\(
0 \le \partial_\mu z^\mu = (\Phi,I\Phi) + \mathcal{O}(\Phi^3)\, .
\)
In the small-\(\Phi\) limit, this implies the inequality
\(
(\Phi,I\Phi)\ge 0
\)
for real \(\Phi\). Furthermore, to quadratic order in \(\Phi\), the time component of the Massieu current reads
\(
z^0=\mathrm{const}-\tfrac12(\Phi,\Phi)\, .
\)
Thus, the bilinear form \((\cdot,\cdot)\) coincides with the entropic quadratic form entering the Onsager reciprocity relations. Applying the Onsager symmetry principle under PT symmetry \cite{GavassinoDistrubingMoving:2026klp}, it follows that the operator \(I\) is real and symmetric. Together with the positivity property derived above, this implies that \(I\) admits a non-negative (Friedrichs \cite[\S 2.3]{TeschlBook}) self-adjoint extension on the whole complex Hilbert space \(\mathfrak{H}\). This proves the first two lines of \eqref{Properties}.

The last statement follows from the conservation of the (quasi)particle current,
\begin{equation}\label{amess}
0=\partial_\mu  J^\mu
= \int \frac{d^D p}{(2\pi)^D}\,(\partial_t+v^j\partial_j)\Phi
= -\int \frac{d^D p}{(2\pi)^D}\, I\Phi
= -\int \frac{d^D p}{(2\pi)^D f_{\mathrm{eq}}}\, f_{\mathrm{eq}}\, I\Phi
= - (f_{\mathrm{eq}},I\Phi)
= - (I f_{\mathrm{eq}},\Phi) \, ,
\end{equation}
where, in the last step, we used the Hermiticity of the operator \(I\).
Since \eqref{amess} must hold for arbitrary local values of \(\Phi\), it follows that
\(I f_{\mathrm{eq}}=0\).
Together with the assumption that no additional linearly independent conserved currents exist, this establishes the final property in \eqref{Properties}.

\bibliography{Biblio}

\label{lastpage}
\end{document}